\documentclass[11pt]{article}

\usepackage{amsmath,amsfonts,bm}

\def\eqref#1{equation~\ref{#1}}

\def\1{\bm{1}}

\DeclareMathAlphabet{\mathsfit}{\encodingdefault}{\sfdefault}{m}{sl}
\SetMathAlphabet{\mathsfit}{bold}{\encodingdefault}{\sfdefault}{bx}{n}

\usepackage[]{acl}

\usepackage{times}
\usepackage{latexsym}
\usepackage[T1]{fontenc}
\usepackage[utf8]{inputenc}
\usepackage{microtype}

\usepackage{url}
\usepackage{graphicx}  
\usepackage{amsfonts}
\usepackage{CJK}
\usepackage{bm}
\usepackage{mathrsfs}
\usepackage{float}
\usepackage{xcolor,colortbl}
\usepackage{adjustbox}
\usepackage{subfigure}
\usepackage{booktabs,multirow}
\usepackage{bigstrut}
\usepackage{paralist}
\usepackage{bbm}
\usepackage{makecell}
\usepackage{nicefrac}       
\usepackage{microtype} 
\usepackage{epsfig}
\usepackage{diagbox}
\usepackage{framed}
\usepackage{empheq}
\usepackage{array}
\usepackage{enumitem}
\usepackage{mdwlist}
\usepackage{xspace,mfirstuc,tabulary}
\usepackage{tcolorbox}
\usepackage{algorithm}
\usepackage{placeins}
\usepackage{algpseudocode}
\usepackage{microtype}

\usepackage[normalem]{ulem}
\useunder{\uline}{\ul}{}
\usepackage{threeparttable}
\usepackage{makecell}
\usepackage{booktabs}

\usepackage{xspace}
\usepackage{natbib}
\usepackage{hyperref}
\usepackage{url}

\usepackage{graphicx}
\usepackage{subfigure}
\usepackage{multirow}
\usepackage{multicol}
\usepackage{pifont}
\usepackage{enumitem}
\usepackage{amsmath}
\usepackage{appendix}

\usepackage{tikz}
\newcommand*{\circled}[1]{\lower.7ex\hbox{\tikz\draw (0pt, 0pt)%
    circle (.5em) node {\makebox[1em][c]{\small #1}};}}

\newcommand{\ie}{{{i.e.,}}\xspace}

\newcommand{\eg}{{{e.g.,}}\xspace}

\newcommand{\tlc}[1]{\makecell[c]{#1}}

\newcommand{\increase}[1]{\textcolor{red}{\scriptsize #1}}

\def\benchmark{{\sc CodeAgentBench}\xspace}
\def\textbfbenchmark{{\sc \textbf{CodeAgentBench}}\xspace}
\def\method{{\sc CodeAgent}\xspace}
\def\textbfmethod{{\sc \textbf{CodeAgent}}\xspace}

\title{\textbfmethod: Enhancing Code Generation with Tool-Integrated Agent Systems for Real-World Repo-level Coding Challenges}

\author{Kechi Zhang
 \\
  Affiliation / Address line 1 \\
  Affiliation / Address line 2 \\
  Affiliation / Address line 3 \\
  \texttt{email@domain} \\\And
  Jia Li \\
  Affiliation / Address line 1 \\
  Affiliation / Address line 2 \\
  Affiliation / Address line 3 \\
  \texttt{email@domain} \\}

  \author{Kechi Zhang\footnotemark[1], \ Jia Li\footnotemark[1], \ Ge Li\footnotemark[2], \ Xianjie Shi, \ Zhi Jin\footnotemark[2] \\
Key Lab of High Confidence Software Technology (PKU), Ministry of Education \\
School of Computer Science, Peking University, China \\
\texttt{\{zhangkechi,lijiaa,lige\}@pku.edu.cn}, \\ \texttt{2100013180@stu.pku.edu.cn}, \\ \texttt{zhijin@pku.edu.cn}}

\begin{document}
\maketitle
\renewcommand{\thefootnote}{\fnsymbol{footnote}}
\footnotetext[1]{The two authors share equal contribution.}
\footnotetext[2]{Corresponding authors.}
\renewcommand{\thefootnote}{\arabic{footnote}}
\begin{abstract}

Large Language Models (LLMs) have shown promise in automated code generation but typically excel only in simpler tasks such as generating standalone code units. However, real-world software development often involves complex code repositories with complex dependencies and extensive documentation. 
To enable LLMs to handle these real-world repo-level code generation, we present \method, a novel LLM-based agent framework that employs external tools for effective repo-level code generation. \method integrates five programming tools, enabling interaction with software artifacts for information retrieval, code implementation, and code testing. We implement four agent strategies to optimize these tools' usage. To the best of our knowledge, \method is the first agent framework specifically for repo-level code generation. 
In order to measure the effectiveness of our method at the repository level, we design a repo-level benchmark \benchmark. The performance on this benchmark shows a significant improvement brought by our method, with improvements in pass rate ranging from 2.0 to 15.8. 
Further tests on the HumanEval benchmark confirm \method's adaptability and efficacy across various code generation tasks. Notably, \method outperforms commercial products like GitHub Copilot, showcasing superior accuracy and efficiency. These results demonstrate \method's robust capabilities in code generation, highlighting its potential for real-world repo-level coding challenges.

\end{abstract}

\nocite{Ando2005,andrew2007scalable,rasooli-tetrault-2015}

\maketitle

\section{Introduction}
Code generation automatically generates programs for the natural language (NL) requirement. Recent years have seen a trend in tackling code generation tasks with large language models (LLMs), such as Code Llama \cite{roziere2023code}, StarCoder \cite{li2023starcoder}, and DeepSeekCoder \cite{DeepSeek}. Many efforts have been performed \cite{zhang2023self, luo2023wizardcoder, zheng2023codegeex} and shown impressive code generation abilities.

Despite achieving satisfactory performances, these studies mainly focus on simple generation scenarios including statement-level and function-level code generation. Statement-level code generation \cite{iyer2018mapping, athiwaratkun2022multi} aims to output statement-specific source codes. Function-level code generation \cite{chen2021evaluating, austin2021program, hendrycks2021measuring} predicts independent code that only invokes built-in functions and APIs from third-party libraries. For both scenarios, the length of the generated code is rather short, and they only generate standalone code units. However, more than 70\% functions in the open-source projects are non-standalone \cite{yu2023codereval}. Developers typically write programs based on specific code environments, generally referring to code repositories. These repo-level code snippets usually have intricate contextual dependencies, which is too complex for existing LLMs to handle and generate \cite{deveval}.

To enhance the efficacy of LLMs in repo-level code generation tasks, we draw inspiration from human programming practices. Developers typically employ a variety of tools to aid in complex programming. For instance, they might utilize search engines to explore key concepts or static analysis tools to identify pre-existing functions or classes. These tools are instrumental in the development of code projects.
Embracing this idea, we propose a novel LLM-based agent framework \method that leverages external tools to help LLMs in repo-level code generation. 
With five programming tools, \method is capable of interacting with the software artifacts, including retrieving useful information, finding existing code symbols in the repository, and handling essential code testing. 
To guide LLMs to efficiently use tools, we draw on four agent strategies covering ReAct, Tool-Planning, OpenAIFunc, and Rule-based form. Based on agent strategies, LLMs can automatically select suitable tools for each repo-level task, finally providing a comprehensive response. 

In order to measure the effectiveness of our method at the code repository, we manually construct \benchmark, a benchmark specifically for repo-level code generation with a total of 101 functions and classes sourced from real code projects. 
It provides rich information about the repository, such as documentation and contextual dependency, to help LLMs better understand it. 
We further conduct extensive experiments for evaluation. We apply \method to nine powerful open-source and closed-source LLMs with parameter sizes ranging from 13B to 175B to show the universality. Compared to directly generating from LLMs, experimental results on \benchmark reveal that \method achieves significant improvements ranging from 2.0 to an extraordinary 15.8 across various LLMs. 
Further evaluations on well-known function-level benchmark HumanEval \cite{chen2021evaluating} confirm \method's versatility in diverse code generation tasks.
Remarkably, when compared to commercial products like GitHub Copilot \cite{dakhel2023github}, \method stands out, demonstrating superior accuracy. These findings highlight the robust practical capabilities of \method in the code generation community, underscoring its potential to evolve real-world repo-level coding challenges. We summarize our main contributions:

\begin{itemize}
\item We make an attempt to investigate repo-level code generation, which has crucial worth for understanding LLMs' performance in practical code generation scenarios.
\item  We propose \method, an LLM-based agent framework for repo-level code generation. It develops five external programming tools to help LLMs complete the whole generation process and draw on four agent strategies to automatically optimize tools’ usage.
\item We construct \benchmark, a repo-level code generation benchmark, which has high-quality code repositories and covers diverse topics. 
\item  Experimental results on nine LLMs show \method's versatility and effectiveness in diverse code generation tasks, highlighting its potential for resolving real-world repo-level coding challenges.
\end{itemize}

\section{Background}
\label{sec:back}

\subsection{LLMs and Agents for Code Generation}

LLMs have shown impressive capabilities in code generation since they have billions of parameters trained on a large amount of corpus with different training objectives. 
Recently, OpenAI \footnote{https://openai.com/} proposes GPT-3.5 and GPT-4 series models (\eg ChatGPT \cite{Chat}), which have shown strong generation abilities in coding. 
There are also various open-soured work, such as CodeGen \cite{nijkamp2022codegen}, StarCoder \cite{li2023starcoder}, Code Llama \cite{roziere2023code}, WizardCoder \cite{luo2023wizardcoder} and DeepSeekCoder \cite{DeepSeek}. 

Recent research has also increasingly shown that LLMs can be instrumental in developing AI agents \cite{agentfoundation,agentsurvey, agentsurvey2,hugginggpt,Gorilla,ToolLLM}. Examples such as ToolFormer \cite{schick2023toolformer}, Auto-GPT \cite{AutoGPT}, BabyAGI \cite{BabyAGI}, KwaiAgents \cite{KwaiAgents} and ToolCoder \cite{zhang2023toolcoder} demonstrate LLMs’ proficiency in tool utilization for complex tasks.
Some studies such as self-edit \cite{zhang2023self} and self-debug \cite{selfdebug} have demonstrated that code models possess the capability for multi-round interaction and repair.
Nowadays, some work has also demonstrated the effectiveness of agent systems in complex code programming tasks, such as OpenDevin \cite{opendevign}, SWE-Agent \cite{sweagent}.
In this paper, we select GPT-4 \cite{GPT-4}, GPT-3.5 \cite{GPT-3.5}, and other powerful LLMs to design coding agent systems for real-world repo-level code generation.

\subsection{Code Generation Tasks} \label{benchmarksss}

Existing code generation tasks mainly focus on generating \textbf{standalone code units}, including statement-level \cite{yin2018learning} and function-level generation \cite{hendrycks2021measuring, chen2021evaluating}.  
The generated programs are usually short and are independent of other codes.
However, in software development, programmers mainly work within a code environment. They extend their functionalities based on the foundational code framework. Inspired by this, some studies \cite{yu2023codereval, liao2023context} introduce intricate programming tasks that are based on particular code environments such as projects and code repositories. Nevertheless, these studies only provide limited constraint information to LLMs, containing the requirements, signature information, and restricted code dependencies, leading to a difference in programming information needs from humans. Some work targets real-world GitHub issues for code model to resolve, such as SWE-bench \cite{swebench}. To get closer to realistic programming scenarios, we formalize the repo-level code generation task and propose \method to help LLMs handle this complex task. We construct a repo-level code generation benchmark \benchmark to evaluate our method and provide an analysis of benchmarks commonly used for these generation tasks in Table \ref{tab:motivating}. Compared with existing code generation tasks, repo-level code generation is more consistent in real-world programming scenarios, fostering the evolvement of the code generation community.

\section{Repo-level Code Generation Task}
\label{sec:task}
\begin{figure*}[t]
\centering
\includegraphics[width=\linewidth]{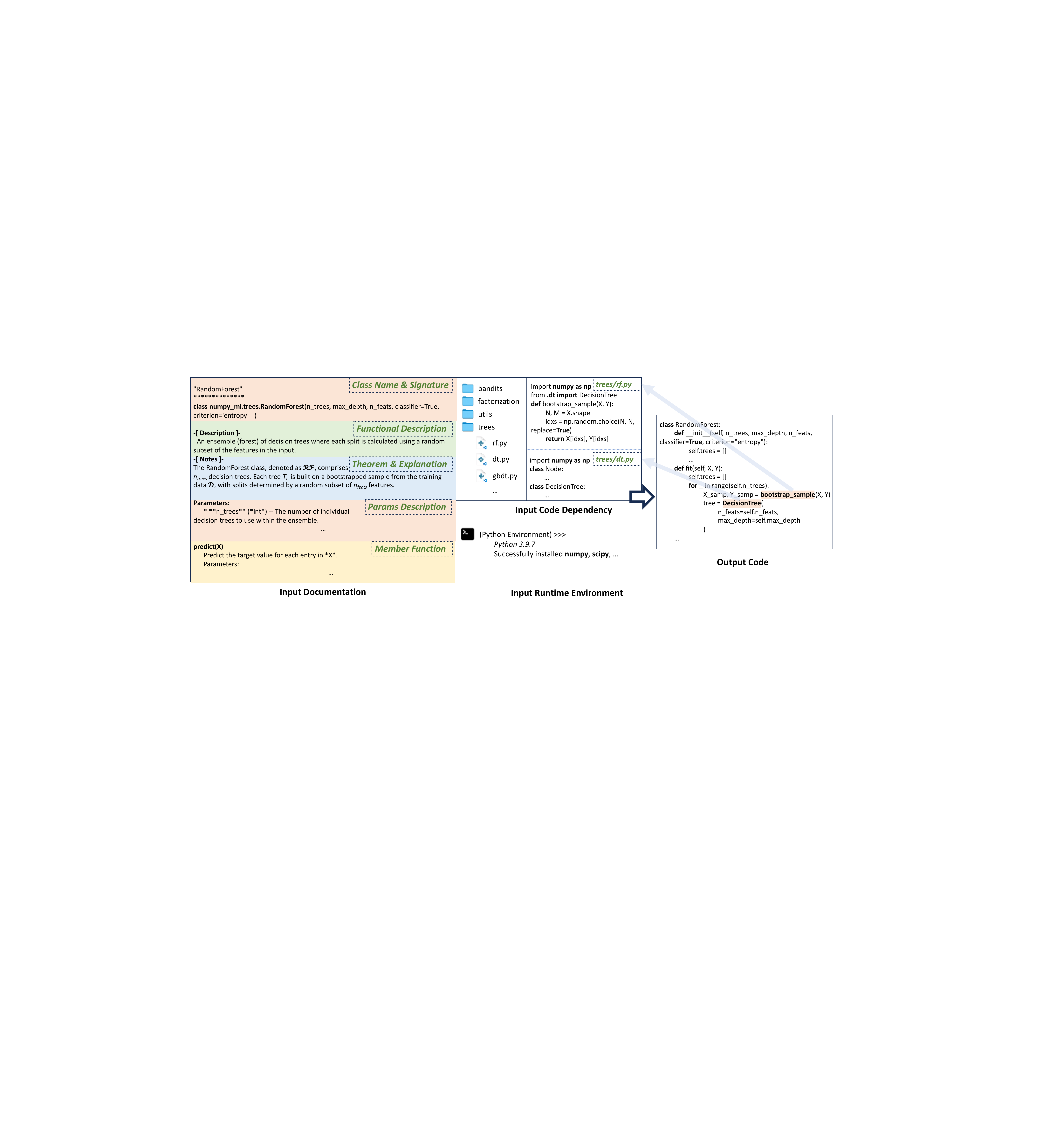}  
\caption{An illustrative example of the repo-level code generation. The task input contains complex descriptions, code dependencies, and runtime environment, which is more realistic than the existing benchmark.}
\label{fig:codeagentbench}
\end{figure*}

To fill the gap between existing code generation tasks and practical coding scenarios, we formalize the repo-level code generation task. Since a code repository generally contains intricate invocation relationships, only with a deep understanding of the code repository can LLMs generate satisfying programs that not only adhere to requirements but also seamlessly integrate with the current repository. 
Given a code repository, the repo-level code generation task aims to generate code based on all the software artifacts included in the repository, encompassing the \textbf{documentation}, \textbf{code dependency}, \textbf{runtime environment}, which form the task input.
Here we give a detailed description of its composition format.
Figure \ref{fig:codeagentbench} shows an illustration of the repo-level code generation task.

\paragraph{Documentation}
It describes the generation targets and is the main input component of repo-level code generation.
The documentation provides additional supporting information beyond the NL requirements. It contains class-level (class name, signature, and member function) and function-level (functional description, and params description) information of targets. Typically, the correctness of generated programs is verified with the test suite. The generated programs must conform to the interface (\eg the input parameters). Thus, the documentation also provides the type and interpretation of input parameters and output values. In addition, considering that requirements usually contain domain-specific terminologies, the documentation explains these terms as well, such as mathematical theorems. 
As shown in Figure \ref{fig:codeagentbench}, documentation of the project contains rich information, where different elements are highlighted with diverse colors. 

\paragraph{Contextual Dependency}

A key distinction of our new task from other independent code generation tasks is its inclusion of contextual dependencies. This aspect is crucial, as classes or functions typically interact with other code segments within the repository, such as import statements or other user-defined classes and functions. These interactions may occur within the same file or across multiple files. For instance, to implement the \textit{RandomForest} class in Figure \ref{fig:codeagentbench}, it is necessary to utilize the \textit{bootstrap\_sample} function from \textit{rf.py} and the \textit{DecisionTree} class from \textit{dt.py}, demonstrating the intricate code contextual dependencies involved.

\paragraph{Runtime Environment}
\label{sec:benchruntime}
Different from natural language, program language is executable. Whether programs return target results after execution is a crucial manner to verify the correctness of generated programs. Developers typically depend on the execution feedback to correct errors in programs. The runtime environment provides all configurations needed to run the code repository and offers convenient interaction to ensure an all-sided evaluation of LLMs' performance on repo-level code generation.

\section{\textbfmethod Method}

\begin{figure}[t]
\centering
  \includegraphics[width=\columnwidth]{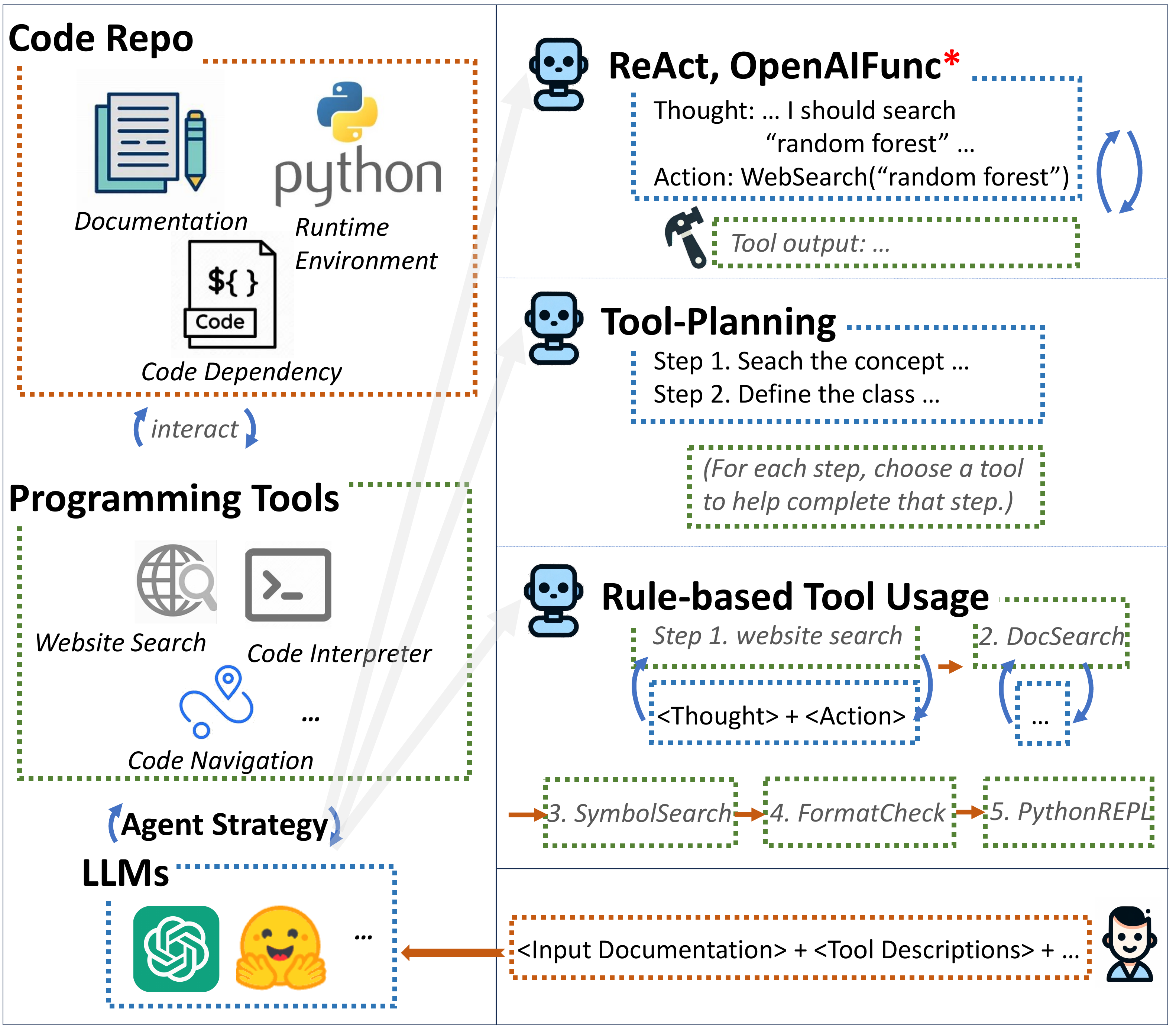}  
\caption{\textbf{\textit{Left}}: Overview of  \method . With our designed programming tools and agent strategies, LLMs interact with code repositories and generate repo-level code. \textbf{\textit{Right}}: Illustration of agent strategies in \method. "\textit{OpenAIFunc}" is similar to "\textit{ReAct}" in the interaction mode, with some differences in the content generated by LLMs and the format of tool callings.}
\label{fig:method}
\label{fig:strategy}
\end{figure}

We introduce a novel LLM-based agent framework \method that leverages external tools to enhance the problem-solving abilities of LLMs in intricate repo-level code generation.
\method seamlessly pauses generation whenever tools are called and resumes generation by integrating their outputs. These tools can assist LLMs with the entire code generation process, including information retrieval, code implementation, and code testing as shown in Table \ref{tab:tools}, thus interacting with the software artifacts (Section \ref{tools}). Providing LLMs with access to tools, \method explores four agent strategies to optimize these tools’ usage (Section \ref{strategy}). 
Figure \ref{fig:method} illustrates the overview of our \method.

\subsection{Designed Programming Tools} \label{tools}

Given a requirement, developers usually first gather relevant knowledge, then find and modify existing programs to meet the requirement, and finally verify programs with the assistance of tools. To mimic this process, we develop several programming tools that are specifically designed for LLMs.
\method incorporates these external tools from three perspectives: information retrieval, code implementation, and code testing, which are commonly used by programmers in their daily work. 

\begin{table}[!t] 
\footnotesize
\setlength{\tabcolsep}{1pt} 
    \centering
    \resizebox{1.0\linewidth}{!}{
\centering
\begin{tabular}{cll}
\toprule
Tool Domain                            & Tool Name              & Usage Pattern                                        \\
\midrule
\multirow{2}{*}{\makecell{Information \\ Retrieval}} & Website Search         &   \textit{WebSearch(input\_query)}        \\ \cmidrule{2-3} 
                                       & Documentation Reading  & \textit{DocSearch(input\_name)}                 \\
                                       \midrule
\makecell{Code \\ Implementation}                    & Code Symbol Navigation & \textit{\makecell{SymbolSearch(module\_path \\or input\_name)}} \\
\midrule
\multirow{2}{*}{Code Testing}          & Format Checker         & \textit{FormatCheck()}                         \\ \cmidrule{2-3} 
                                       & Code Interpreter       & \textit{PythonREPL(input\_code)}               \\
                                       \bottomrule
\end{tabular}
}
\caption{Programming tool statistics in \method}
   \label{tab:tools}
\end{table}

\subsubsection{Information Retrieval Tools}

Information retrieval tools are responsible for analyzing repositories and collecting resources, which is pivotal in understanding the problem domain. We develop popular website search and documentation reading as information retrieval tools.

\paragraph{Website Search} Programmers often share solutions for various programming problems on websites where search engines consider them as knowledge resources. When encountering similar problems, developers only submit a question query to a search engine. The engine can provide useful programming suggestions. Inspired by this, \method uses a popular search engine DuckDuckGo\footnote{https://duckduckgo.com/} to choose the most relevant websites, and then apply LLMs to summarize the website content as the final tool output \footnote{We choose \textit{DuckDuckGo} because it provides a cheaper and more convenient API than other search engines such as \textit{Google} and \textit{Bing}.}. In the process, we block websites that may lead to data leakage. The usage pattern of this tool is formatted as: \textit{WebSearch(input\_query)}, which will return the formatted content searched from websites. 

\paragraph{Documentation Reading} Besides gathering information from websites, we also retrieve relevant knowledge from the documentation of the repository. To achieve this, \method leverages BM25 \cite{robertson2009probabilistic} as the documentation reading tool. Given a class name or function name, it can retrieve correlative content from the documentation as its output. If the result is too long, the tool will use the LLM to summarize it and then provide it to LLMs for code generation. This tool is designed in the format: \textit{DocSearch(input\_name)}. 

\subsubsection{Code Implementation Tools}

Code implementation tools aim to provide relevant code items (\ie pre-defined symbol names and code snippets) in the code repository. LLMs modify and integrate these items into the generation process.
It not only expedites the development process but also encourages code reuse. We build a code symbol navigation tool to help LLMs implement code snippets.

\paragraph{Code Symbol Navigation} We use \textit{tree-sitter} \footnote{https://tree-sitter.github.io/tree-sitter/} to design the code symbol navigation tool. This tool explores code items from two types. The first type is oriented to the file or module-oriented parsing, where the tool performs static analysis of a file or module and provides symbol names defined in it, encompassing global variables, function names, and class names. The other type is the class or function symbol navigation. Given a class or function name, the tool finds its definition from the code repository. Combining the two types, this tool can traverse predefined source code within a repository, empowering LLMs to understand intricate dependencies and reuse codes. This tool is designed in the format: \textit{SymbolSearch(module\_path or input\_name)}. The tool will detect what the input is and return the corresponding results (\eg all defined symbols in the given file path or the implementation code corresponding to the given symbol name). When no parameters are provided, the default value is the path of the current file.

\subsubsection{Code Testing Tools}

After acquiring generated codes, we design code testing tools to format and test them, enhancing their correctness and readability. 

\paragraph{Format Checker} The tool is built to check the format correctness of generated codes. Specifically, we develop \textit{Black} \footnote{https://github.com/psf/black} as the format checker. It can check format errors such as indentation misalignment and missing keywords. Subsequently, it tries to rectify these errors and reorganizes code statements, enhancing the correctness and readability of generated codes. The usage pattern of this tool is: \textit{FormatCheck()}, which will automatically format the most recently generated code and return the formatted version.

\paragraph{Code Interpreter} The tool focuses on examining the syntax and function of programs. It furnishes a runtime environment so that LLMs can debug generated codes with execution feedback. The tool requires LLMs to provide a program to be executed, and then runs the code in the repository environment. Meanwhile, LLMs generate some test cases to verify whether the output of the generated code meets the expected results. When occurring errors, this tool will offer error information to facilitate LLMs to fix bugs until programs are error-free, which has been proven to be effective by many existing works \cite{chen2022codet, zhang2023self} to correct output programs. The runtime environment is prepared for each task, as described in Section \ref{sec:benchruntime}. 
This tool is designed in the format: \textit{PythonREPL(input\_code)}, and the tool will return the executed result of the input code.

\subsection{Agent Strategy} \label{strategy}

To guide LLMs to leverage these powerful tools properly, we develop four agent strategies for repo-level code generation, including ReAct, Tool-Planning, OpenAIFunc, and Rule-based Tool Usage. 
The interaction between LLMs and external tools is based on LangChain \footnote{https://python.langchain.com}.

\paragraph{ReAct} This strategy \cite{yao2022react} prompts LLMs to generate reasoning traces and task-related actions in an interlaced fashion. Based on actions, ReAct selects the proper external tools and invokes them by providing input. The strategy then treats the output of tools as additional knowledge and decides whether to generate a final code or invoke other tools for further processing.

\paragraph{Tool-Planning} We propose a variant, \ie Tool-Planning,  of Planning strategy \cite{wang2023plan} that makes a plan before solving problems and has shown effectiveness in many studies \cite{zhang2022automatic, jiang2023self}. Different from Planning, our strategy can invoke proper tools based on the plan. Specifically, Tool-Planning first makes a plan to divide an entire task into several subtasks and then performs subtasks according to the plan. For complex subtasks, it will automatically choose an appropriate tool to assist LLMs in code generation.

\paragraph{OpenAIFunc} 
Recently, some models (\eg GPT-3.5 \cite{GPT-3.5} and GPT-4 \cite{GPT-4}) have the function-calling ability provided by OpenAI \cite{OpenAIFunc}. The interaction mode is similar to that of "ReAct", with some differences in the content generated by LLMs and the format of calling external tools.

\paragraph{Rule-based Tool Usage}

When faced with a complex problem, programmers often first learn related knowledge, then write programs, and check the function of programs. Inspired by the workflow, we propose a rule-based strategy. 

This strategy defines the order of tool usage and interlinks these tools by prompts. I) LLMs leverage website search to gather useful online information; II) LLMs then use documentation reading tool to search relevant classes and functions; III) Code symbol navigation is required to select and view the source codes of related classes and functions. Based on the above information, LLMs generate programs; IV) Subsequently, LLMs invoke the format checker to check the syntax and format of generated programs; V) Finally, LLMs use the code interpreter to evaluate the functional correctness of programs. Based on the feedback information, LLMs fix errors within programs. 
For each part, LLMs will autonomously cycle through the use of tools until it decides to move on to the next part or the cycle reaches its limit number (\eg 3).

\section{Experiment}

We perform extensive experiments to answer three research questions:
(1) How much can \method improve the advanced code generation LLMs on repo-level code generation (Section \ref{sec:codeagentbenchrst});
(2) What is the improvement of our \method on classical code generation such as HumanEval (Section \ref{sec:humanevalrst});
(3) To what extent do our selected tools in the agent system help for repo-level coding (Section \ref{sec:ablationrst}).

\subsection{Experimental Setup}

\paragraph{Benchmarks}
\label{app:construction}


\begin{table}[!t] 
\footnotesize
    \centering
    \resizebox{\linewidth}{!}{
\centering
\begin{tabular}{lcccc}
\toprule
Name  & Domain       & Samples & \# Line & \# DEP \\
\midrule
numpyml-easy & Machine Learning       & 22            &      10.9         &      0.3           \\
numpyml-hard   & Machine Learning       & 35            &      85.4          &      2.6                    \\
container      & Data Structure         & 4             &      130.3           &   8.0                       \\
micawber       & Information Extraction & 7             &      19.7         &      4.3                 \\
tinydb         & Database               & 21            &       36.7          &     2.7                   \\
websockets     & Networking             & 12            &       91.6             &   7.5                    \\
\midrule
Total &                        & 101           &     57.0          &      3.1        \\
\bottomrule
\end{tabular}
}
\caption{Statistics of \benchmark. \# Line: average lines of code. \# DEP: average number of code dependencies. }
   \label{tab:repobenchmark}
\end{table}

To evaluate our method on repo-level code generation, we follow the format described in Section \ref{sec:task} and construct a new benchmark \textbfbenchmark.
To make \benchmark diverse, we select five prevalent topics judged by ten developers and choose repositories with high stars from GitHub. The selected topics contain machine learning, data structure, information extraction, database, and networking. To ensure the quality, we only select repositories that use \textit{pytest} \footnote{https://docs.pytest.org/} and \textit{unittest}\footnote{https://docs.python.org/3/library/unittest.html} as the test framework and its documentation is generated by \textit{Sphinx}\footnote{https://www.sphinx-doc.org/} tool. 
For writing standards of these test cases, since we opted for projects utilizing the pytest and unittest frameworks, these frameworks ensure consistency in these testing codes. (for example, the pytest framework requires all test functions to have "test\_" as a prefix in their function names and provides uniform guidelines for test assertions).
We also filter out complex repositories that are hard to deploy and test. Then, we extract all functions and classes in code repositories and arrange two participants to sequentially execute them. 
Our construction costs approximately 600 person-hours. Each participant possesses 2-5 years of Python programming experience. Finally, we get 101 functions and classes collected from real code projects in Python. The statistics of \benchmark are shown in Table \ref{tab:repobenchmark}.

The final \benchmark contains 101 samples, and for each task, LLMs are provided with documentation containing the requirements needed to be implemented, along with a set of tools we designed, as well as full access permissions to code files in the repository. We use the self-contained test suite in each code repository to evaluate the correctness of generated programs.

\begin{figure}[t]
\centering
  \includegraphics[width=\columnwidth]{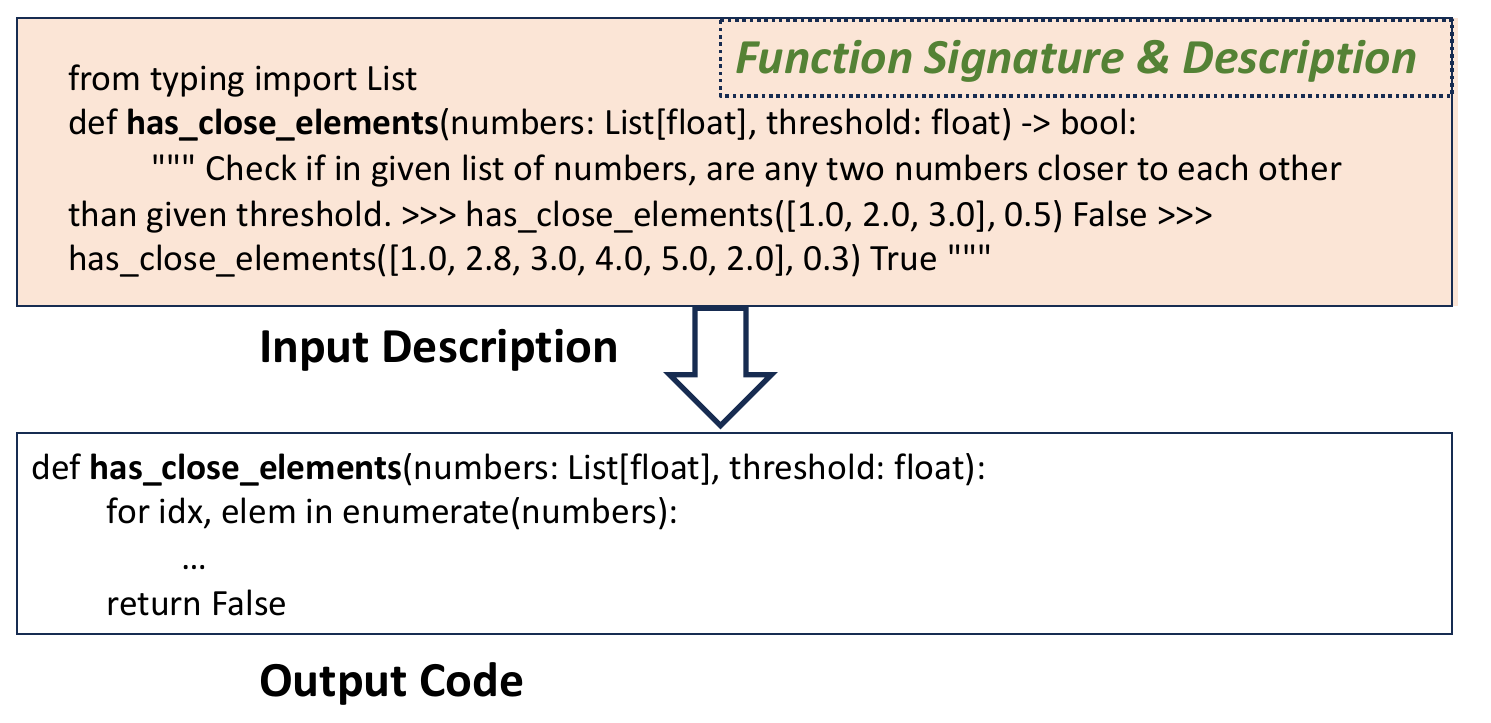}  
\caption{An illustrative example of existing benchmark HumanEval.}
\label{fig:humaneval}
\end{figure}

In addition, to evaluate the generalization ability of \method, we also perform experiments on function-level code generation. In this paper, we use a widely-used function-level benchmark \textbf{HumanEval} \cite{chen2021evaluating}. It contains 164 programming problems with the function signature, docstring, body, and unit tests. In Figure \ref{fig:humaneval}, we give an illustrative example of HumanEval.

\paragraph{Base LLMs} We apply \method to nine most powerful LLMs, including GPT-3-davinci \cite{GPT-3}, GPT-3.5-turbo \cite{GPT-3.5}, GPT-4-turbo \cite{GPT-4}, Claude-2 \cite{Claude}, Llama2-70B-chat \cite{Llama}, Code Llama-34B \cite{roziere2023code}, WizardCoder-34B \cite{luo2023wizardcoder}, DeepSeek-33B \cite{DeepSeek} and Vicuna-13B \cite{chiang2023vicuna}.  Additional descriptions are provided as a part of Table \ref{tab:codeagentbench}.

\paragraph{Metrics}

Following previous works \cite{zan2022cert, zheng2023codegeex}, we use the pass rate as the metric, where we treat the generated program correctly only if its output is consistent with all ground truths of the test suite.  Specifically, we are mainly concerned with \textbf{Pass@1} \cite{chen2021evaluating}, which is a representative of the Pass@k family, because in real-world scenarios, we usually only consider the single generated code.

\subsection{Repo-level Coding Performance}
\label{sec:codeagentbenchrst}

\begin{table*}[!t] 
\footnotesize
    \centering
    \resizebox{\linewidth}{!}{
\centering
\begin{tabular}{l|c|
>{\columncolor[HTML]{DEE0E3}}c| cccc}
\toprule
\textit{\textbf{Models}} & Scales & \textbf{NoAgent} & \textbf{Rule-based} & \textbf{ReAct} & \textbf{Tool-Planning} & \textbf{OpenAIFunc} \\
\midrule
\textbf{\textit{Closed source LLM}} & &  &  & & & \\
GPT-3-davinci \cite{GPT-3}    & 175B  & 16.8 & \textbf{24.8} (\increase{ $\uparrow$ 7.9})  & 22.8 (\increase{ $\uparrow$ 5.9}) & 18.8 (\increase{ $\uparrow$ 2.1}) & -  \\
GPT-3.5-turbo \cite{GPT-3.5}           & -                  & 19.8       & \textbf{31.7}  (\increase{ $\uparrow$ 11.9})          & 30.7 (\increase{ $\uparrow$ 10.8})    & 21.8 (\increase{ $\uparrow$ 2.0})   & 28.7 (\increase{ $\uparrow$ 8.9})         \\
GPT-4-turbo  \cite{GPT-4}           &  -               & 21.8       & \textbf{37.6} (\increase{ $\uparrow$ 15.8})         & 34.7 (\increase{ $\uparrow$ 12.9})    & 25.7 (\increase{ $\uparrow$ 4.0})   & 34.7 (\increase{ $\uparrow$ 12.9})         \\
Claude-2  \cite{Claude}         &  -              &  8.9                &    \textbf{10.9} (\increase{ $\uparrow$ 2.0})                 &    9.9 (\increase{ $\uparrow$ 1.0})            &     9.9 (\increase{ $\uparrow$ 1.0})        &    -               \\
\midrule
\textbf{\textit{Open source LLM}} & &  &  & & & \\
Llama2-70B-chat  \cite{Llama}    & 70B            &     10.9        &    \textbf{12.9} (\increase{ $\uparrow$ 2.0})         &  11.9 (\increase{ $\uparrow$ 1.1})               &  11.9 (\increase{ $\uparrow$ 1.1})       &        -          \\
Code Llama-34B \cite{roziere2023code}          & 34B            &   2.0         &     \textbf{5.0} (\increase{ $\uparrow$ 3.0})            &   4.0 (\increase{ $\uparrow$ 2.0})        &    4.0  (\increase{ $\uparrow$ 2.0})     &  -                \\
WizardCoder-34B \cite{luo2023wizardcoder}       & 34B             &    2.0          &  \textbf{6.9} (\increase{ $\uparrow$ 5.0})           &    5.0 (\increase{ $\uparrow$ 2.7})        &    4.0 (\increase{ $\uparrow$ 2.0})      &     -           \\
DeepSeek-33B  \cite{DeepSeek}       & 33B            &      13.9            &   \textbf{24.8} (\increase{ $\uparrow$ 10.9})             &   20.8 (\increase{ $\uparrow$ 6.9})       &    15.8 (\increase{ $\uparrow$ 2.0})     &    -              \\
Vicuna-13B  \cite{chiang2023vicuna}          & 13B            &  1.0         & \textbf{1.0}           &      0.0          &    0.0         &     -             \\
\bottomrule
\end{tabular}
}
\caption{The Pass@1 results of different agent strategies on \benchmark. {``NoAgent''} refers to the baseline where LLMs generate code solely based on the provided documentation.} 
   \label{tab:codeagentbench}
\end{table*}

In our experiments, we utilized our specially designed repo-level benchmark, \benchmark, to assess the efficacy of \method in enhancing the performance of nine prominent code LLMs. The results are presented in Table \ref{tab:codeagentbench}.  

Our proposed \benchmark proves to be substantially more challenging than existing benchmarks, as evidenced by the relatively lower pass rates. On all base LLMs with various sizes, \method consistently delivers significant performance improvements. Specifically, for GPT-4 model \cite{GPT-4}, we observe a maximum increase of 15.8, equating to a 72.7\% relative enhancement over the baseline, \ie NoAgent. The improvements of other LLMs range from 2.0 to an impressive 15.8, underscoring the effectiveness of our proposed approach. This demonstrates that the tools integrated within \method provide useful information, aiding LLMs in producing accurate code solutions and effectively tackling complex repo-level coding challenges.

Across different LLMs, a notable trend is that more advanced LLMs exhibit greater improvements with the application of \method. However, for Vicuna-13B model \cite{chiang2023vicuna}, performance on \benchmark is notably poor, showing no appreciable enhancement with the agent strategy. In contrast, the improvement is quite pronounced for other high-capacity LLMs. Furthermore, we find that different agent strategies yield varying levels of enhancement. Among these strategies, Rule-based and ReAct strategies are more effective, whereas Tool-Plannig strategy appears less suited for the task.

\subsection{Function-level Coding Performance}
\label{sec:humanevalrst}

\begin{table*}[!t] 
\footnotesize
    \centering
\centering
\begin{tabular}{l|
>{\columncolor[HTML]{DEE0E3}}c| cccc}
\toprule
\textit{\textbf{Models}} & \textbf{NoAgent} & \textbf{Rule-based} & \textbf{ReAct} & \textbf{Plan} & \textbf{OpenAIFunc} \\
\midrule
GPT-3.5-turbo \cite{GPT-3.5}                   & 72.6       & \textbf{82.3}  (\increase{ $\uparrow$ 9.7})          & 79.3 (\increase{ $\uparrow$ 6.7})    & 73.8 (\increase{ $\uparrow$ 1.2})   & 81.1 (\increase{ $\uparrow$ 8.5})         \\
CodeLLaMA-34B \cite{roziere2023code}                   &   51.8         &     \textbf{59.7} (\increase{ $\uparrow$ 7.9})            &   58.2 (\increase{ $\uparrow$ 6.4})        &    54.1  (\increase{ $\uparrow$ 2.3})     &  -                \\
WizardCoder-34B \cite{luo2023wizardcoder}                  &    73.2          & \textbf{79.4} (\increase{ $\uparrow$ 6.2})           &    77.6 (\increase{ $\uparrow$ 4.4})        &    75.6 (\increase{ $\uparrow$ 2.4})      &     -           \\
DeepSeek-33B  \cite{DeepSeek}                &     78.7   & \textbf{84.8} (\increase{ $\uparrow$ 6.1})             &   83.5 (\increase{ $\uparrow$ 4.8})       &    81.1 (\increase{ $\uparrow$ 2.4})     &    -              \\
\bottomrule
\end{tabular}
\caption{The Pass@1 results of different agent strategies on the HumanEval benchmark.}
   \label{tab:humaneval}
\end{table*}

We further apply our \method to function-level code generation with the well-known HumanEval benchmark \cite{chen2021evaluating}. We adapt our approach to this scenario by omitting the documentation reading tool and code symbol navigation. The adjustment is necessitated as these tools are not applicable to the standalone code generation task. For this task, we strategically selected a range of representative LLMs for evaluation, constrained by our available resources and computational capacity. The pass rate results are detailed in Table \ref{tab:humaneval}.

The results once again highlight the efficacy of \method in enhancing the performance of code LLMs across all metrics. Notably, the maximum improvements observed for each model span from 6.1 to 9.7 on Pass@1.
These findings underscore the versatility and effectiveness of our \method in augmenting the capabilities of LLMs across a variety of code generation tasks.

\subsection{Ablation Study}
\label{sec:ablationrst}

\begin{table}[!t] 
\footnotesize
    \centering
    \resizebox{\linewidth}{!}{
\centering
\begin{tabular}{l|c|c}
\toprule
                & \textbf{\# Usage} & \textbf{Ablation Result} \\
                \midrule
\textit{GPT-3.5-ReAct}    & -           & 30.7             \\
\midrule
\textit{Website Search}       & 0.30         & 27.7 (\increase{ $\downarrow$ 3.0})            \\
\textit{Documentation Reading}      & 0.84        & 26.7   (\increase{ $\downarrow$ 4.0})         \\
\textit{Code Symbol Navigation}         & 2.45        & 22.8    (\increase{ $\downarrow$ 7.9})        \\
\textit{Format Check}   & 0.17        & 29.7    (\increase{ $\downarrow$ 1.0})        \\
\textit{Code Interpreter} & 0.22        & 29.7   (\increase{ $\downarrow$ 1.0})         \\
\midrule
\textit{GPT-3.5-NoAgent}  & -           & 19.8           \\
\bottomrule
\end{tabular}
}
\caption{Average tool usage number and ablation result on \benchmark for GPT-3.5-ReAct.}
   \label{tab:toolablation}
\end{table}

To investigate the influence of tools incorporated in \method, we conduct an ablation study focusing on tool utilization in repo-level code generation.
We choose GPT-3.5-turbo with ReAct as the base model, named GPT-3.5-ReAct. We meticulously track the usage frequency of each tool during code generation processes, with the statistics presented in Table \ref{tab:toolablation} under the column \textit{\# Usage}.
Subsequently, we exclude one tool at a time from our approach, allowing us to isolate and understand the individual contribution of each tool. The performances of these ablation scenarios are shown in Table \ref{tab:toolablation}, categorized under the column \textit{Ablation Result}.

Our findings reveal that the code symbol navigation tool is particularly pivotal in our agent system. On average, \method utilizes this tool approximately 2.45 times per code generation, a frequency higher than the counterpart of other tools. Notably, the performance significantly declines when this tool is omitted, underscoring its critical role in enhancing the effectiveness of our approach. 
Furthermore, the ablation results confirm that each tool in our agent system contributes positively to the overall improvement. This evidence not only validates the effectiveness of our strategy design but also highlights the utility of programming tools in addressing the repo-level coding task.

\section{Discussion}

\subsection{Compared with Commercial Products}

\label{sec:comparedcopilot}

\begin{table}[!t] 
\footnotesize
    \centering
    \resizebox{0.95\linewidth}{!}{
\centering
\begin{tabular}{l|cc}
\toprule
\textbf{}                       & NumpyML-easy & NumpyML-hard \\
\midrule
\textit{\textbf{Our Agent}} &              &              \\
GPT-3.5                          & \textbf{14}           & \textbf{3}            \\
GPT-4                            & \textbf{17}           & \textbf{5}            \\
\midrule
\textit{\textbf{IDE Product}} &              &              \\
GitHub Copilot                         & 7          &   1         \\
Amazon CodeWhisperer                   & 5           &  0          \\
\midrule
\textit{\textbf{Agent Product}}         &              &              \\
AutoGPT (with GPT-4)             & 2            & 0           \\
\bottomrule
\end{tabular}
}
\caption{Performance compared with commercial programming products (the number of solved problems).}
   \label{tab:copilot}
\end{table}

Nowadays, a lot of mature commercial products are available to support complex code generation tasks. 
It is essential to compare \method with these established products. 
We categorize them into two distinct groups:
(1) \textit{IDE Products} are AI-powered autocomplete-style suggestion tools integrated within IDE software. Notable examples are \textit{GitHub Copilot} \cite{copilotgithub} and \textit{Amazon CodeWhisperer} \cite{CodeWhisperer}.
(2) \textit{Agent Products} encompass autonomous agents driven by GPT-4 \cite{GPT-4}. They are capable of executing a variety of tasks, including coding, such as well-known \textit{AutoGPT} \cite{AutoGPT}.

Considering that IDE products are primarily designed as completion systems, we limit human interactions to less than three times per task to ensure a fair comparison. The evaluation is conducted on the \textit{numpyml} subset of \benchmark manually by an experienced Python developer. Table \ref{tab:copilot} shows the number of solved problems on different products and our \method.

The results demonstrate that \method works better than existing products on complex coding scenarios. In addition, despite both \method and AutoGPT being agent-based approaches, \method exhibits numerous optimizations tailored for repo-level coding tasks, thereby making it better than AutoGPT in the task. 
Compared to IDE products that can also analyze complex code dependencies, our method benefits from the flexibility inherent in the agent system, resulting in a substantial lead over IDE products. 

\subsection{Qualitative Analysis}
\label{sec:case}

We explore generated cases to assess \method (\eg GPT-3.5-ReAct) and the baseline model (\eg GPT-3.5-NoAgent). The comparative analysis is shown in Figure \ref{fig:case1} and Figure \ref{fig:case2}.

\method typically begins with examining the code dependencies in the repository, subsequently refining its code generation strategy through a step-by-step process known as ``chain-of-thought''. 
As in Figure \ref{fig:case1}, the input documentation specifies the need for a class with member functions \textit{set\_params} and \textit{summary}. \method, assisting with the symbol navigation tool, finds the base class and identifies the member function \textit{\_kernel} as a key component for implementation. This is reflected in the generated thought process:
\begin{quote}
    \textit{"The set\_params and summary methods can be inherited from the base class without modifications ... The `\_kernel' method needs to be overridden ..."}
    
    \textit{\footnotesize (Generated by \method-\textit{GPT-3.5-ReAct})}
\end{quote}
On the contrary, GPT-3.5-NoAgent lacks access to detailed information on code structures, resulting in incorrect code solutions, as depicted in Figure \ref{fig:case2}.

\section{Conclusion}
We formalize the repo-level code generation task to evolve real-world coding challenges. To enhance LLMs to handle repo-level code generation, we propose \method, a novel LLM-based agent framework. \method develops five programming tools, enabling LLMs to interact with software artifacts, and designs four agent strategies to optimize tools' usage. To evaluate the effectiveness of our \method, we construct \benchmark, a new benchmark for repo-level code generation that includes rich information about the code repository. Experiments on nine LLMs show that \method achieves a significant improvement on diverse programming tasks, highlighting its potential in real-world coding challenges.

\section{Acknowledgments}
This research is supported by the National Natural Science Foundation of China under Grant No.62192733, 61832009, 62192731, 62192730, 62072007, the Major Program (JD) of Hubei Province (No.2023BAA024).


\section*{Limitation}
Although our work is a very early exploration of this area, there are several limitations on our work that we aim to address as quickly as possible:

Firstly, we propose a new task format for the repo-level code generation task and release \benchmark. Our preliminary experiments prove that the impact of LLMs' memorization on pre-training data is slight for fair evaluation. However, it still needs further experiments to eliminate this hidden danger. We will follow the relevant research to further understand its influence on our proposed benchmark.

Secondly, we only incorporate simple tools to \method. Some advanced programming tools are not explored. The limitation may restrict the agent’s ability in some challenging scenarios. 

Thirdly, in Section \ref{sec:comparedcopilot}, the comparison with commercial products is not rigorous since
experiments are done manually. We will study how to evaluate IDE products more standardly.

Finally, since LLMs are very sensitive to input prompts, it is very important to optimize prompts in the agent system. We will continue to explore better agent strategies based on the current approach.

\section*{Ethics Consideration}
CodeAgent and its benchmark are inspired and collected from real-world code repositories. We manually check all samples in our benchmark. We ensure all samples do not contain private information or offensive content. Throughout our experiments, we diligently annotated the sources of all used data, ensuring compliance with the respective license specifications.

\bibliography{main}


\clearpage
\newpage
\appendix

\begin{figure*}[!h]
\centering
  \includegraphics[width=1.88\columnwidth]{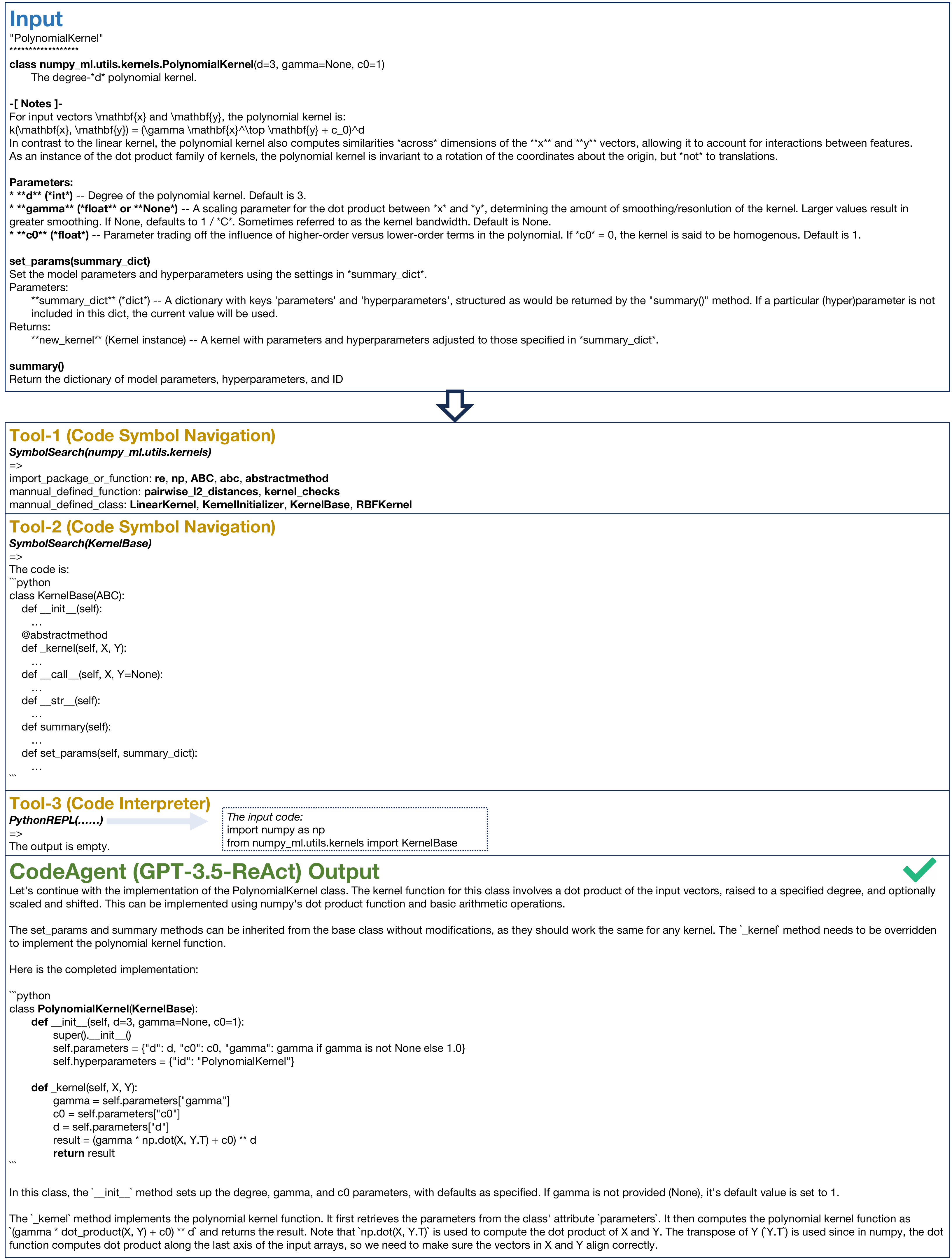}  
\caption{Case study of our proposed \method (GPT-3.5-ReAct) on \benchmark. We omit some input prompts and LLM's thoughts for viewing convenience.}
\label{fig:case1}
\end{figure*}

\begin{figure*}[!h]
\centering
  \includegraphics[width=1.88\columnwidth]{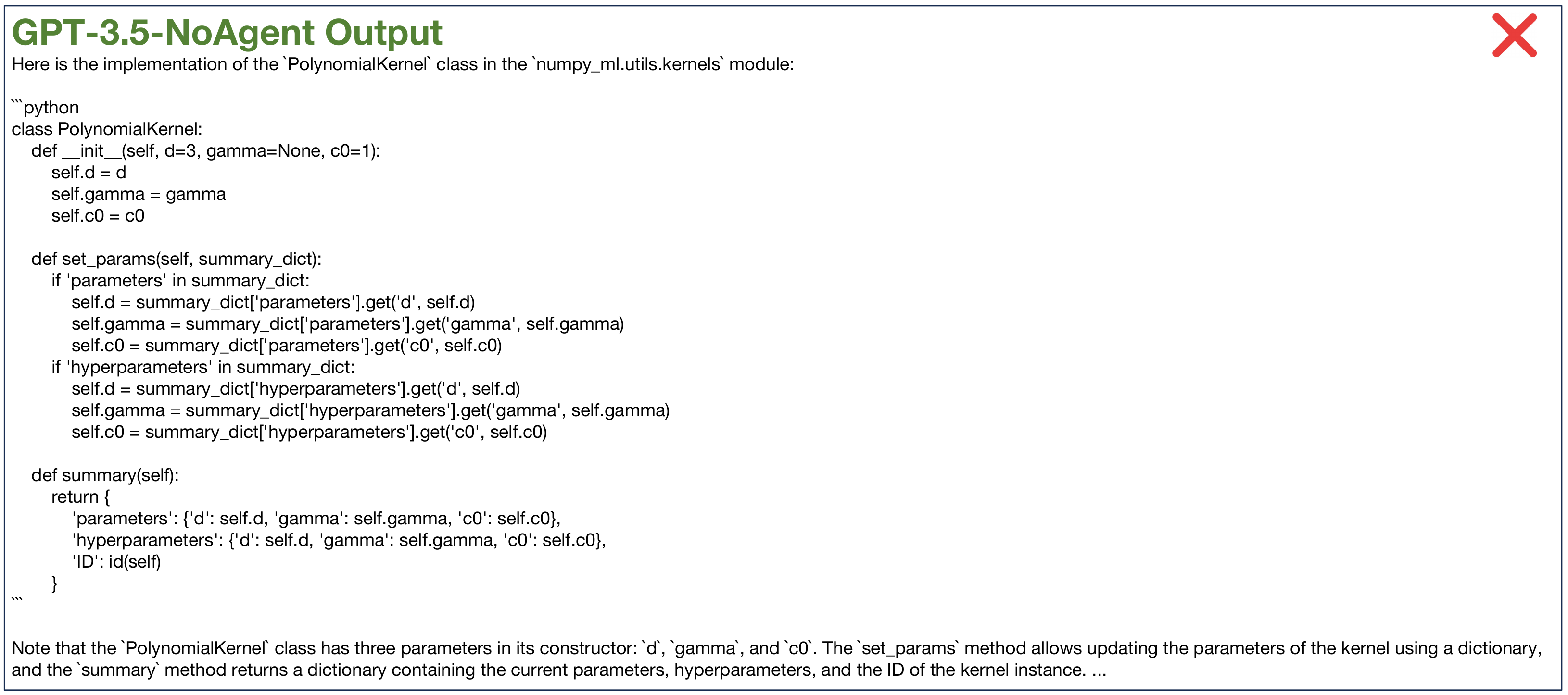}  
\caption{Case study of the baseline GPT-3.5-NoAgent on \benchmark. Compared with Figure \ref{fig:case1}, GPT-3.5-NoAgent generates the incorrect programs.}
\label{fig:case2}
\end{figure*}

\section{Details of Case Study}
Here we show the illustration of the case study for \method (GPT-3.5-ReAct) and GPT-3.5-NoAgent in Figures \ref{fig:case1} and \ref{fig:case2}. 

We can find a distinct operational pattern in \method in Figure \ref{fig:case1}. 
Through meticulous analysis, \method leverages code symbol navigation tool to scrutinize information within the `utils.kernels' module, where the target class for implementation resides. 
Our custom-designed tool proficiently navigates to the module, offering insights into its contents, including package details, defined functions and classes, through a static analysis process. 
Importantly, \method discovers a crucial class named `KernelBase' and obtains detailed information about it with another use of the tool. Within `KernelBase', there is an abstract method named `\textit{\_kernel}' that needs to be implemented. \method recognizes this method as essential for the development process, highlighting its importance.
Compared with the NoAgent in Figure \ref{fig:case2}, our approach accurately captures this content hidden in the complex information in the code repository, and precisely implements the final code.

We also notice that during the third tool invocation, \method calls the code interpreter tool and execute a piece of code that appears insignificant. 
We have observed similar situations in other cases as well. We attribute this to LLMs still lacking proficient mastery of some complex programming tools. 
This insight directs our future research towards enhancing LLMs' ability to more effectively use complex programming tools.

    

\section{Details of \textbfbenchmark}

\label{codeagentbench}

In this section, we introduce the details of our \benchmark benchmark. We describe its composition format (Section \ref{app:format}), the construction process (Section \ref{app:construction}), and provide a detailed comparison with existing benchmarks (Section \ref{app:comparison}).

\subsection{Benchmark Composition}
\label{app:format}

Code repository contains intricate invocation relationships. Only with a deep understanding of code repository can LLMs generate satisfying programs that not only adhere to requirements but also seamlessly integrate with the current repository. Inspired by this, each task of our benchmark provides rich information, encompassing the documentation, code dependency, runtime environment, self-contained test suite, and canonical solution, which form the input and output.

\subsubsection{Benchmark Input}
\paragraph{Documentation}
Documentations are the main input component of our benchmark and describe the generation targets.
We follow the code documentation format used in a popular documentation creation tool Sphinx \footnote{https://www.sphinx-doc.org/}. Figure \ref{fig:codeagentbench} illustrates an example of documentation in \benchmark, where different elements are highlighted with diverse colors. When accomplishing a new task, our prepared documentation can provide LLMs with all-sided details that need to be considered to ensure that the generation target has been well-defined and constrained.

\paragraph{Contextual Dependency}


Contextual dependency is an important role in our benchmark.
To accurately identify these dependencies, we developed a static analysis tool using \textit{tree-sitter} \footnote{https://tree-sitter.github.io/tree-sitter/}. Our designed tool allows us to extract all user-defined elements (such as class names, function names, constants, and global variables) and public library names from each file. These elements are then stored in a knowledge base. For any given function, we use this knowledge base to locate its source file, parse the file to identify all user-defined symbols and public libraries, and finally determine its contextual dependencies by exact matching of symbol names and scopes. On average, each sample in \benchmark involves around 3.1 code dependencies, thereby closely simulating real-world programming conditions. Detailed information is shown in Table \ref{tab:repobenchmark}.

\paragraph{Runtime Environment}
Developers often use feedback from running programs to find and fix mistakes. In \benchmark, we build a sandbox environment for each task. The sandbox environment provides all configurations needed to run the repository and offers convenient interaction to ensure an all-sided evaluation of LLMs' performance on repo-level code generation.

\subsubsection{Benchmark Ground-truth Output}
\paragraph{Canonical Solution} We use the answers included in the repository as the initial solutions and invite three participants to manually refine them. The first participant checks surface errors of solutions based on the repository information. The second person runs the solutions to identify and fix execution bugs. The last participant is responsible for executing solutions with the test suite, aiming to ensure its functional correctness. Through the iterative process, we can ensure the robustness and reliability of solutions as much as possible.  

\subsubsection{Benchmark Evaluation}
\paragraph{Self-Contained Test Suite} 
To evaluate the correctness of generated programs, \benchmark furnishes a self-contained test suite for each task. We first analyze and extract test cases contained in the repository. We then invite two participants to manually add test cases to enhance its coverage as much as possible. In \benchmark, each task has at least one unit test case. Whereafter, another participant manually checks the correctness of the test suite. Given a new task, we run the corresponding unit test code to verify the generated programs based on our sandbox environment. We treat the generated program correctly only if its output aligns with all ground truths of the test suite. For fairness, LLMs can not access the test suite during code generation.

\subsection{Compared with Existing Benchmarks}
\label{app:comparison}

\begin{table*}[!h] 
\footnotesize
    \centering
    
    \resizebox{\linewidth}{!}{
\centering
\begin{tabular}{lcccccccc}
\toprule

Benchmark  & Language  & Source & Task   & Samples    & \# Tests & \# Line &\# Tokens & \# Input  \\
\midrule
\midrule
CoNaLA \cite{yin2018learning}      & Python    & Stack Overflow  & Statement-level & 500   & \ding{54}   &1  & 4.6       & NL  \\
\midrule
Concode ~\cite{iyer2018mapping}      & Java    & Github   & Function-level  & 2000      & \ding{54}  & -  & 26.3    & NL      \\
\midrule
APPS \cite{hendrycks2021measuring}        & Python  & Contest Sites  & Competitive & 5000   & \ding{52}    &21.4  & 58    & \tlc{NL + IO}                              \\
\midrule
HumanEval \cite{chen2021evaluating}   & Python   & Manual   & Function-level   & 164   & \ding{52} &11.5 & 24.4  & \tlc{NL + SIG + IO}         \\
\midrule
MBXP  \cite{athiwaratkun2022multi}      & Multilingual       & Manual   & Function-level  & 974  & \ding{52} &6.8 &24.2    & NL      \\
\midrule
InterCode  \cite{yang2023intercode}      & SQL, Bash    & Manual    & Function-level  & 200, 1034 & \ding{52}  & - & -    & \tlc{NL + ENV }                                               \\
\midrule
CodeContests \cite{li2022competition} & Python, C++  & Contest Sites & Competitive & 165   & \ding{52}   & 59.8 &184.8   & \tlc{NL + IO}                              \\
\midrule
ClassEval  \cite{du2023classeval}       & Python       & Manual  & Class-level  & 100    & \ding{52}   & 45.7 & 123.7   & \tlc{NL + CLA}                                        \\
\midrule
CoderEval \cite{yu2023codereval}   & Python, Java & Github  & Project-level  & 230   & \ding{52}  &30.0 & 108.2  & \tlc{NL + SIG}       \\
\midrule
RepoEval \cite{liao2023context}         & Python       & Github   & Repository-level   & 383    & \ding{54}   & -  & -  & \tlc{NL + SIG} 
\\
\midrule
\benchmark               & Python       & Github  &Repository-level       & 101       & \ding{52}   & 57.0 &  477.6  & \tlc{Software Artifacts \\ (NL + DOC \\ + DEP + ENV)} \\
\bottomrule
\end{tabular}
}
\caption{The statistics of existing widely-used code generation benchmarks. \# Tests:  whether a benchmark has the test suite. \# Line: average lines of code. \# Tokens: average number of tokens. \# Input: Input information of LLMs. NL: Natural language requirement. IO: Input and output pairs. SIG: Function signature. CLA: Class skeleton as described in Section \ref{benchmarksss}. ENV: Runtime environment. DOC: Code documentation. DEP: Code dependency. }
   \label{tab:motivating}
\end{table*}

We perform a detailed analysis of existing code generation benchmarks in Table \ref{tab:motivating}. 
Compared to the previous benchmarks, our \benchmark has two main advantages. On the one hand, it is closer to real-world code generation scenarios. On the other hand, \benchmark provides pretty complex information that is related to the code repository, including documentation, contextual dependency, runtime environments, and test suites.

\end{document}